\documentclass[aps,prd,reprint,nofootinbib,superscriptaddress]{revtex4-1}
\usepackage[utf8]{inputenc}
\usepackage[english]{babel}
\usepackage[fleqn]{amsmath}
\usepackage{mathtools,amssymb,bm}
\usepackage[retain-unity-mantissa = false]{siunitx}
\usepackage{slashed}
\usepackage{enumitem}
\usepackage{notes2bib}
\usepackage{color}

\newcommand{\diff}{\mathrm d}
\newcommand*{\eh}[1]{\mathrm e^{#1}}
\renewcommand{\vec}[1]{{\bm{#1}}}

\newcommand{\eperp}{\epsilon_\perp}
\let\Texp\relax
\DeclareMathOperator{\Texp}{Texp}
\begin{document}
{
\makeatletter
\let\cat@comma@active\@empty
\makeatother
\title{Afterglow of the dynamical Schwinger process: soft photons amass}
\author{A.~Otto}
\author{B.~K\"ampfer}
\affiliation{
Institute of Radiation Physics, Helmholtz-Zentrum Dresden-Rossendorf, 01328
Dresden, Germany\\
Institut f\"ur Theoretische Physik, Technische Universit\"at Dresden, 01068
Dresden, Germany}

\date{\today}

\begin{abstract}
We consider the conversion of an electric field into photons as a secondary
probe of the dynamical Schwinger process.
In spatially homogeneous electric fields, quantum fluctuations of
electron-positron ($e^+e^-$) pairs are lifted on the mass shell leaving
asymptotically a small finite pair density.
The $e^+e^-$ dynamics in turn couples to the quantized photon field and drives
its on-shell mode occupation.
The spectral properties of the emerging asymptotic photons accompanying the
Schwinger process are calculated in lowest-order perturbation theory.
Soft photons in the optical range are produced amass in the sub critical region,
thus providing a promising discovery avenue, e.g.\ for laser parameters of the
Extreme Light Initiative (ELI-NP) to be put in operation soon.
\end{abstract}

\maketitle
}
\section{Introduction}
The Schwinger process refers to lifting virtual pair fluctuations on the mass
shell by a suitable external field.
Considering electron-positron ($e^+e^-$) pairs, Schwinger~\cite{schwinger}
evaluated within the Quantum ElectroDynamics (QED) approach the decay of the
vacuum under the impact of an electric background field, thus formalizing the
pioneering investigations of Sauter~\cite{sauter}.
The history of this interesting branch of strong-field physics and its modern
developments are reviewed in~\cite{gelis_schwinger_2015}, where also many
relevant citations can be found.
By now, a multitude of scenarios has been explored, where such a pair (or,
generically, particle) creation mechanism is of utmost importance.
Examples include Hawking radiation~\cite{hawking_radiation,
gibbons_cosmological_1977}, Unruh radiation~\cite{unruh_notes_1976},
cosmological particle production~\cite{parker_particle_1968,
*parker_quantized_1969,*parker_quantized_1971} and hadron production from
chromoelectric flux tubes~\cite{casher_chromoelectric-flux-tube_1979}.
Focusing on the electromagnetic -- that is QED -- sector of the standard model
of particle physics, much hope is put on the rapidly evolving technology of
ultra-high-intense laser facilities~\cite{di_piazza_extremely_2012} to achieve
in the future electric field strengths sufficiently large to get a direct
experimental access to $e^+e^-$ pairs ``created from vacuum''.
Various field models have been considered which could provide a route towards a
detection of such pairs, among them the superposition of differently shaped
laser fields~\cite{orthaber_momentum_2011,kohlfurst_optimizing_2013,
hebenstreit_optimization_2014,akal_electron-positron_2014,
otto_lifting_2015,otto_dynamical_2015}.
Since the plain Schwinger process yield in a spatially homogeneous electric
field is $\propto\exp(-\pi E_c/E_0)$ with $E_c = m^2/e = \SI{1.3e16}{V/cm}$ (for
electrons and positrons with mass $m$ and charges $\mp e$ in natural units), the
presently attainable fields $E_0\ll E_c$ can yield only exceedingly small
numbers~\cite{ringwald_pair_2001} due to the small tunneling probability.
Spatial inhomogeneities further diminish the pair abundancies~\cite
{ilderton_nonperturbative_2015}, up to a critical suppression~\cite
{gies_critical_2016}.
One option is therefore to elucidate, whether secondary probes are suitable to
identify the pair creation.
This is the motivation of the present paper:
We consider real photon production accompanying the pair creation process.
Similar to the McLerran-Toimela formula~\cite{mclerran_photon_1985}, which is
widely used for evaluating the photon emissivity of the thermalized quark-gluon
plasma, we restrict ourselves on the leading-order $e^2$ yield at
asymptotically large times where a clear particle--anti-particle definition is
applicable.
(To emphasize the asymptotic character of the calculated photon spectrum we
consider here the time-limited action of the background field.)
Clearly, the $e^+e^-$ fluctuation dynamics regarding the \emph{out}-state is
distinctively different from a plasma dynamics, even when accounting for thermal
off-equilibrium effects~\cite{shen_photon_2015}.
Despite of this, but similar to a (nearly) thermalized plasma, our system
facilitates the emission of real photons of all wavelengths, with details
depending on the background field dynamics.

A different, in some sense opposite (similar to the relation of Breit-Wheeler
pair production and Schwinger pair production), approach is followed in~\cite
{karbstein_stimulated_2015}:
Photon production is considered as scattering off the vacuum as a consequence of
the interaction of several, e.g.\ three, incoming real photon beams with a
vacuum loop.
The impact of the frequency composition of the newly created photons is markedly
different and to be contrasted with our continous spectral distribution emerging
off the spatially extended system.
The process considered in~\cite{karbstein_stimulated_2015} refers to an exclusive 1-photon \emph{out}-state, while
we have in mind the inclusive 1-photon spectrum due to the above mentioned --
very restricted -- analogy to a plasma-like system.

The analogy to a radiating plasma system has been utilized, e.g.\ in~\cite{
blaschke_dynamical_2009}, as evidenced by a kinetic theory formula for $2\to2$
processes with on-shell particles and the folding of two distribution functions
by the $e^+e^-\to2\gamma$ cross section.
Another approach is persued in~\cite{kuchiev_enhanced_2015} where recollisions
of once produced $e^+e^-$ lead to hard photons, again via the $e^+e^-\to2\gamma$
cross section.
This is to be contrasted with~\cite{blaschke_self-consistent_2010}, where the
evolution of the photon correlation function is considered, formulated as
leading order in the BBKGY hierachy, which -- after employing some truncation
and diagonalization -- results in a kinetic equation similar to that in the
$e^+e^-$ sector.
The authors of~\cite{blaschke_self-consistent_2010} find a soft photon spectrum
inversely proportional to the photon frequency and proportional to total
$e^+e^-$ number, quite different from our result presented below, which predicts
a large number of photons in the optical regime, thus overcoming the unfavorably
small number of residual (and hence hardly measurable) $e^+e^-$ pairs at present
and near-future laser installations.

Our paper is organized as follows.
In section~II, we present a formula for calculating the photon spectrum which
arises, in first-order perturbation theory, as a consequence of Schwinger pair
production.
Based on such an approach to the time-integrated final-stage photon yield (the
``afterglow'') we provide in section~III numerical evaluations for the Sauter
pulse and a periodic pulse modulated by a time-limited envelope as important
examples of field configurations, which have been also employed formerly in
studying the plain Schwinger pair production.
Here, we exemplify furthermore that the superposition of external fields with
different time scales can result in order-of-magnitude amplifications of the
emergent photon yield, similarly to the dynamically assisted Schwinger process.
Our summary can be found in section~IV.
This main body of the paper uncovers the phenomenological aspects of our
approach, up to an estimate of an ELI-NP-related prediction.
All formal aspects of our approach are relegated to the appendices.
Appendix~A spells out in detail the foundations of our photon spectrum formula
by exploiting suitable transits between Heisenberg picture and interaction
picture to arrive at a solution to the photon wave equation and its relation to
the fermion dynamics.
Appendix~B discusses the soft-photon spectrum and recalls the Bogoliubov
transformation which is needed to make relevant formulas for fermion dynamics
explicitly.

\section{A formula for the photon spectrum}
The impact of an external electric field on the quantum vacuum consists in
inducing a vacuum current which in turn is a source of real-photon fluctuations.
In the QED sector, the remainder of the vacuum current is a finite -- and in
general non-trivial -- $e^+e^-$ pair distribution, referring to the Schwinger
process.
We calculate the spectrum of emerging photons by solving the quantized Maxwell
wave equation in first-order perturbation theory as
\begin{align}
&\mkern-3mu f_\gamma(\vec k) = \frac{e^2}{(2\pi)^6} \frac{1}{2\omega}
\int\!\!\diff^3p\sum_{\lambda,r,s}
\left|\epsilon_\lambda^\mu(\vec k)C_{rs\mu}(\vec p,\vec k)\right|^2
\label{n_from_C},\\
&\mkern-3mu C_{rs\mu}(\vec p,\vec k) = \lim_{\varepsilon\to0}
\int_{-\infty}^\infty\!\!\diff t\,f_\varepsilon(t)\label{C}\\
&\mkern120mu\times \bar v_r(t,-\vec p) \gamma_\mu u_s(t,\vec p-\vec k)
\eh{-i\omega t}\nonumber
\end{align}
highlighting the time-asymptotic photon yield and valid for a spatially
homogeneous system.
The photons propagate on the light cone, i.e.\ the frequency $\omega$ and
wave three-vector $\vec k$ are related by $\omega^2-\vec k^2=0$ and their
polarization four-vector $\epsilon_\lambda$ is orthogonal to the wave
four-vector;
$\lambda=1,2$ counts the polarization states.
$f_\varepsilon=\eh{-\varepsilon|t|}$ is an adiabatic switch-on/switch-off
function of the external field, and $\bar v_r$ and $u_r$ are the time dependent
Dirac wave functions in that field.
The details of formal operations to arrive at~(\ref{n_from_C},\ref{C}) are
spelled out in Appendix A.
Equations~(\ref{n_from_C},\ref{C}) allow for the first time a systematic study
of the photon emission accompanying the Schwinger process.
For instance, one can show (see Appendix~B) that the soft photons are
insensitive to details of the transient Fermion dynamics encoded in $u_r$ and
$v_r$, instead they reflect essentially the difference of \emph{in}- and
\emph{out}-vacua.
In contrast, the hard photons do resolve the actual background field dynamics,
albeit in a time-integrated manner.
Here, we meet severe interferences of the various contributions to the time
integral in~\eqref{C}.
In lacking analytical expressions for $\omega\gg m$ we resort to numerical
solutions pointing to an exponential shape.

\section{Numerical results}
\subsection{Sauter pulse}
\begin{figure}
\centering
\includegraphics[width=0.45\textwidth]{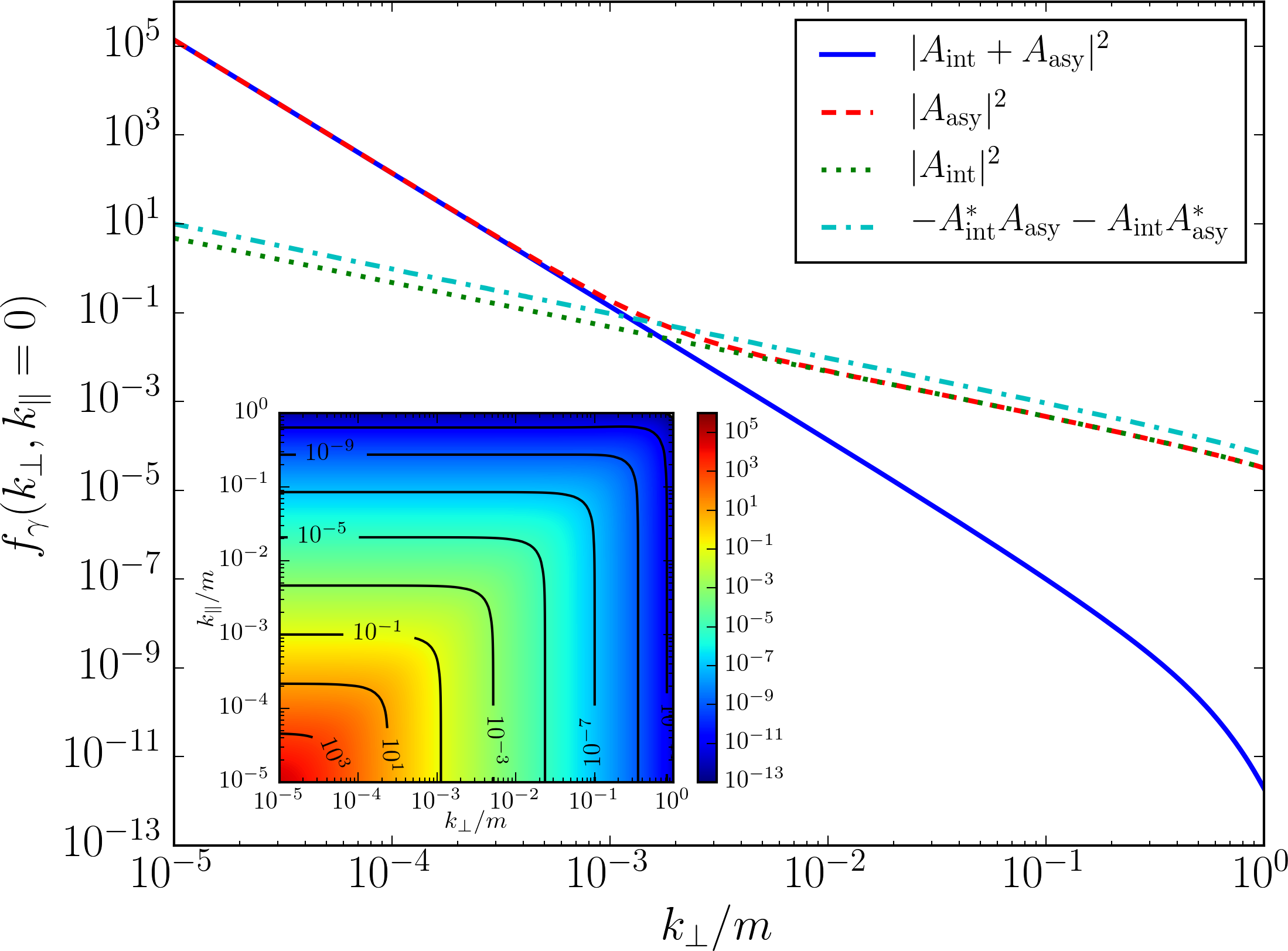}
\caption{Asymptotic phase-space distribution $f_\gamma(\vec k)$ displayed as a
function of $k_\perp$ at $k_\parallel=0$ for the Sauter pulse with $E_0=0.2E_c$
and $\tau=2/m$.
Solid blue curve: full result;
dashed red curve: contribution from the asymptotic time integral in~\eqref{C}
with $A_\text{asy}\propto(\int_{-\infty}^{-t_m} + \int_{t_m}^\infty)\diff t$;
dotted green curve: contribution from the intermediate time integral
$A_\text{int}\propto\int_{-t_m}^{t_m}\diff t$;
dash-dotted cyan curve: the interference term of $A_\text{asy}$ and
$A_\text{int}$;
$A_\text{asy,int}$ are insensitive to variations of the matching time around
$t_m=20\tau$.
The inset exhibits the contour plot of the phase-space distribution
$f_\gamma(k_\perp,k_\parallel)$.}
\label{sauter_cut}
\end{figure}
The Sauter pulse with electric field $E(t) = E_0/\cosh^2(t/\tau)$ and potential
$A(t) = E_0\tau(1+\tanh(t/\tau))$ is an often used external field model which
has an analytical solution of the time evolution of the $e^+e^-$ pair density
$N_{e^+e^-}(t)$~\cite{hebenstreit_optimization_2014};
for $\tau>50/m$ it recovers the seminal Schwinger result.
Even if only $N_{e^+e^-}(t\to\pm\infty)$ has a sensible interpretation in terms
of \emph{in} and \emph{out} asymptotic particle and anti-particle states, a
curious fact is that the mode occupation in an adiabatic basis displays
$N_{e^+e^-}(t\approx0) \ggg N_{e^+e^-}(t\to\infty)$ for deep-subcritical fields
$E_0\ll E_c$~\cite{otto_pair_2016,panferov_assisted_2016}.
Our main result~(\ref{n_from_C},\ref{C}) does not allow to address such an issue.
Instead, we exhibit in Fig.~\ref{sauter_cut} an example of an asymptotic photon
spectrum for parameters $E_0$ and $\tau$ in the subcritical region, $E_0<E_c$,
$\tau>1/m$.
The individual contributions $\left(\int_{-\infty}^{-t_m}
+ \int_{t_m}^\infty\right)\diff t$ and $\int_{-t_m}^{t_m}\diff t$ to the
coefficient~\eqref{C} are separately displayed as a function of $k_\perp$ (the
component of $\vec k$ perpendicular to $\vec E = (0,0,E(t))$) at $k_\parallel=0$
(the component of $\vec k$ parallel to $\vec E$; the full
$k_\perp$-$k_\parallel$ distribution is exhibited in the inset as a contour
plot).
Clearly visible are (i) the  $1/\omega^3$ shape of the soft-photon distribution
and (ii) the onset of the exponential decline of hard photons.
In the optical--UV range, e.g.\ $\omega \sim \num{1e-5} m$, we see a large
phase-space occupancy of $f_\gamma=\num{1e5}$.
(In the spirit of the infrared catastrophe, the number of unobservably soft
photons diverges logarithmically, while the energy emitted per unit volume
remains finite.)
\begin{figure}
\centering
\includegraphics[width=0.45\textwidth]{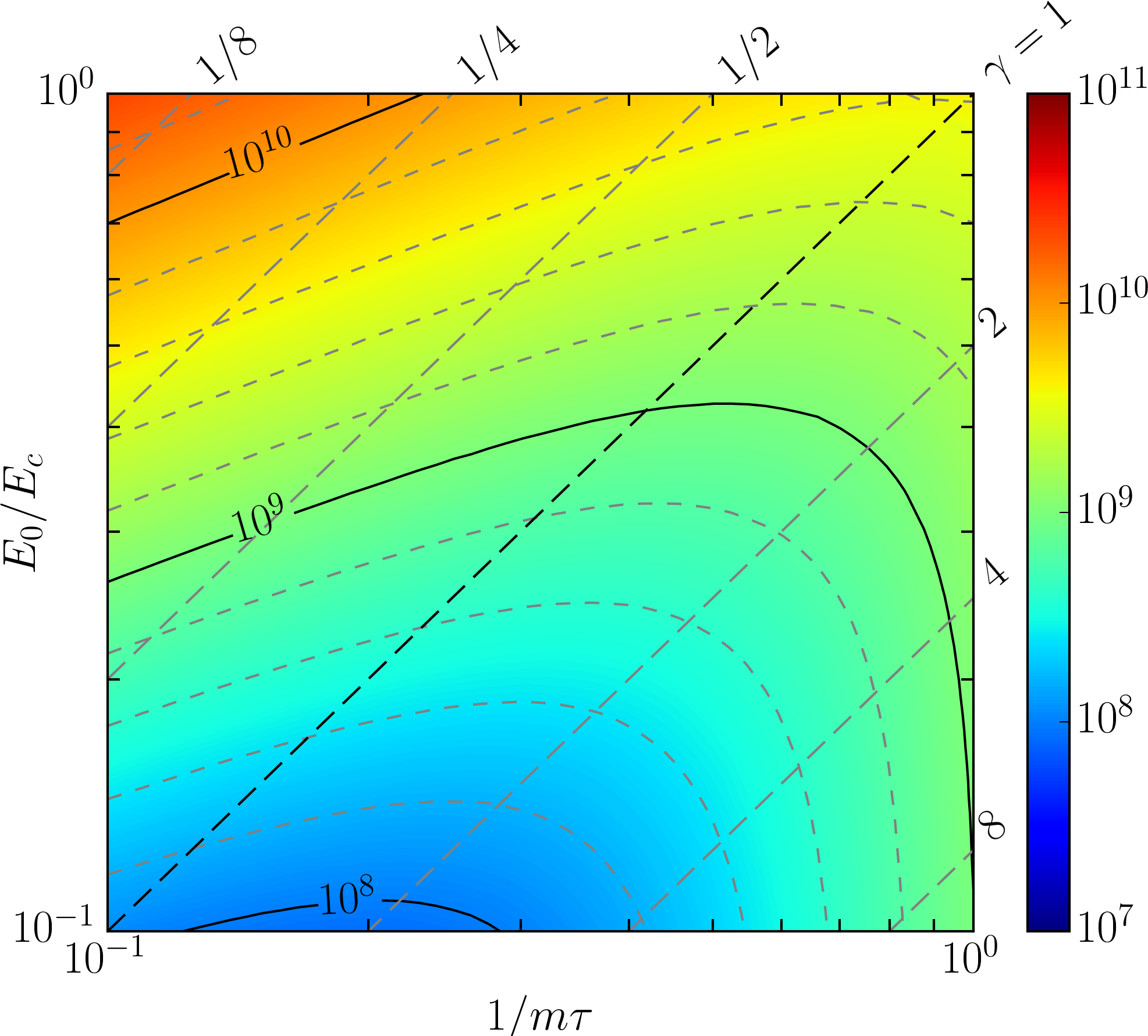}
\caption{Contour plot of the asymptotic photon phase-space occupancy
$f_\gamma(k_\perp=\omega,k_\parallel=0)$ for $\omega=\num{1e-5}m$ normalized to
the asymptotic $e^+e^-$ phase-space occupancy $f_{e^+e^-} = \diff^6N_{e^+e^-}/
\diff^3x\diff^3p$ at $\vec p=0$ for the Sauter pulse.
The diagonal dashed lines display loci of constant Keldysh parameters $\gamma=
\frac{E_c}{E_0}\frac{1}{m\tau}$.
In the tunneling regime, $\gamma<1$.
}
\label{photons_over_pairs}
\end{figure}

Figure~\ref{photons_over_pairs} exhibits the photon numbers at $\omega=\num
{1e-5}m$ normalized to the residual $e^+e^-$ pair number at $\vec p=0$.
Remarkably w.r.t.\ an experimental verification, the soft-photon numbers exceed
by far the residual pair numbers\footnote{%
Reference~\cite{di_piazza_quantum_2010} provides an important example of
particles in an intense external field which emit also multiple photons -- even
hard ones.
}
in the displayed patch of parameter space, e.g.\ $f_\gamma(\omega=\num{1e-5}m)
/f_{e^+e^-}(\vec p=0)=\num{3.2e8}$ at $E_0=0.2E_c$ and $m\tau=2$.
(Due to the $\omega^{-3}$ scaling of $f_\gamma$ for $\omega<0.1m$, one can
deduce from Fig.~\ref{photons_over_pairs} the distribution of other soft-photon
frequencies.)
While encouraging for a detection of the Schwinger process by a
secondary probe, we see a monotonous reduction of the soft photon number
relative to the pair number upon decreasing values of $E_0$, when keeping the
dynamical time scale $\tau$ fixed.
However, extrapolating results of Fig.~\ref{photons_over_pairs} to the regime of
the Nuclear Physics pillar of the Extreme Light Initiative (ELI-NP)\cite
{eli-np}, $E_0=\num{1e-3}E_c$, $\tau=\num{5e5}/m$~\cite{otto_lifting_2015}, the
ratio $f_\gamma(\omega=\num{1e-5}m)/f_{e^+e^-}(\vec p=0)$ becomes favourably
$\num{1e4}$ since both $E_0$ and $\tau$ are diminished.
The employed values of $E_0$ and $\tau$ are deduced from the two-\SI{10}{PW}
laser configuration as core of ELI-NP which is, according to the delivery plan
(cf.~\cite{eli-np}), envisaged to become operational in 2018.
To extrapolate we exploit the apparent relation $\log f_\gamma = a\log E_0/E_c +
b\log m\tau +c$ valid for small Keldysh parameters $\gamma\ll1$.
Therefore, real photons in the optical range, together with their nearly
isentropic radiation pattern (see inset of Fig.~\ref{sauter_cut}) are identified
as promising signature of the Schwinger effect.
Their yield can be enhanced further by multi-scale field configurations.

\subsection{Superposition of fields with different time scales}
The superposition of a strong, slowly varying field with a weaker, fast-varying
field is known to yield a residual pair number which can considerably exceed the
residual pair number of each field alone -- this is the dynamically assisted
Schwinger effect~\cite{schutzhold_dynamically_2008} or assisted dynamical
Schwinger effect~\cite{otto_dynamical_2015}.
Reference~\cite{ilderton_nonperturbative_2015} states in more general terms that
an increasing time-like inhomogeneity of a background field enhances the pair
production.
\begin{figure}
\centering
\includegraphics[width=0.45\textwidth]{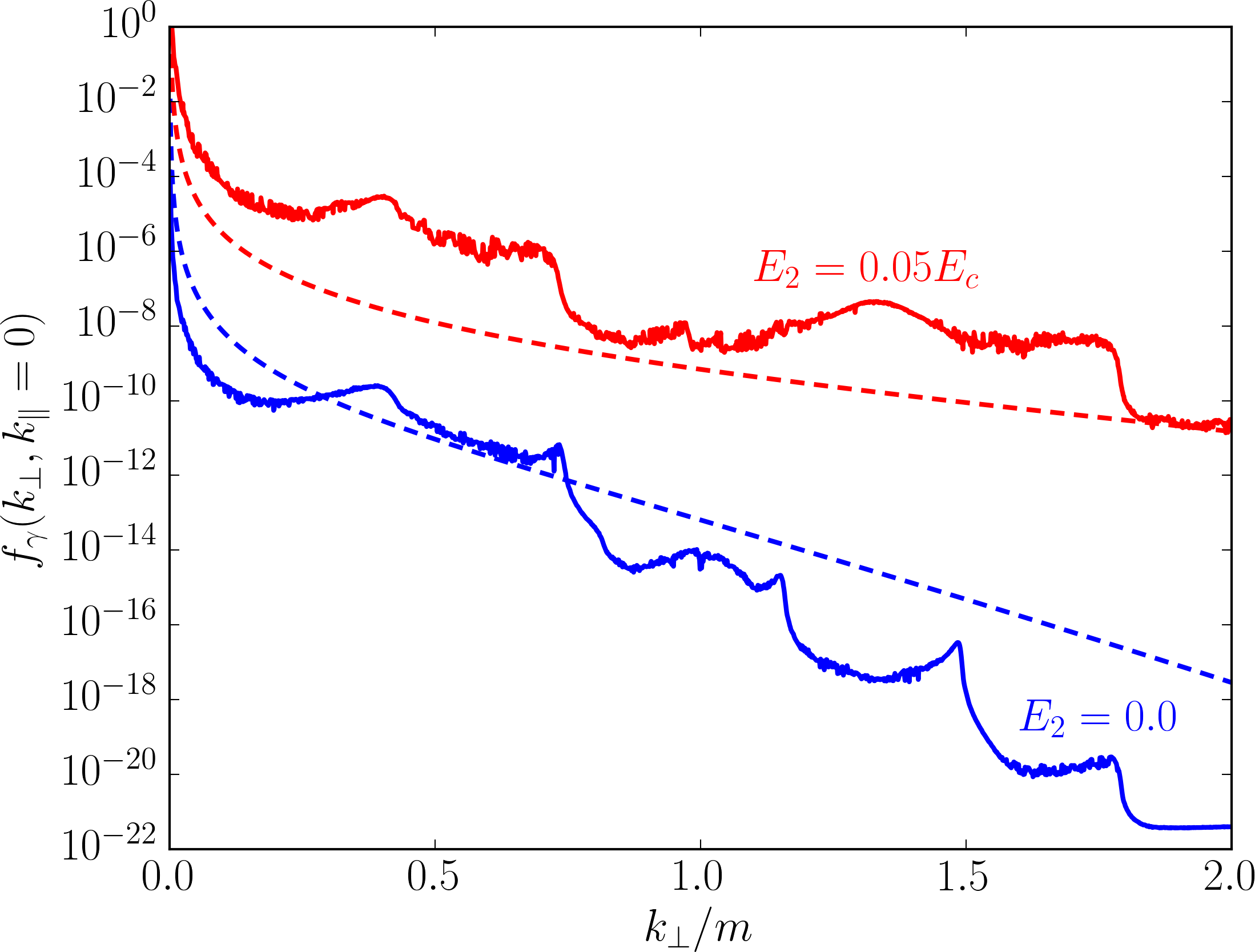}
\caption{Asymptotic photon phase-space occupancy $f_\gamma(\vec k)$ as a
function of $k_\perp$ at $k_\parallel=0$ for the superposition~\protect\eqref
{sauter} of Sauter pulses (dashed curves) and an oscillating field according to
\protect\eqref{sinus} (solid curves) with an
envelope $K(t)$ according to~\cite{otto_lifting_2015} (flat-top interval
$50\cdot2\pi\cdot\tau$ and (de)ramping time(s) $5\cdot2\pi\cdot\tau$).
Parameters are $E_1=0.1E_c$, $\tau=2/m$ and (i) $E_2=0$ (lower blue curves) and
(ii) $E_2=0.05E_c$ and $N=4$ (upper red curves).
Note the exponential shape for hard photons with $\omega>0.5m$ created in the
Sauter pulse.}
\label{sinus_cut}
\end{figure}
Figure~\ref{sinus_cut} unravels an analog effect for the photons when
considering the field model
\begin{align}
E(t) = E_1/\cosh^2(t/\tau) + E_2/\cosh^2(Nt/\tau)\:.
\label{sauter}
\end{align}
Being aware of the rather schematic character of the Sauter pulses employed
above, we include here a field model which may be realized in the anti-nodes of
pairwise counter propagating linearly polarized (laser) photon beams resulting
in a purely electric background field $E(t)$ with potential $A(t)$ when ignoring
the magnetic field components and the spatial inhomogeneity outside the
anti-nodes.
To be specific, our field model is
\begin{align}
E(t) = K(t)\{E_1\sin(t/\tau) + E_2\sin(Nt/\tau)\},
\label{sinus}
\end{align}
where $K(t)$ is a $C^\infty$ smooth envelope function in~\cite
{otto_lifting_2015}.
In both cases, the Sauter pulse~\eqref{sauter} and the model~\eqref{sinus}, the
increased temporal inhomogeneity amplifies significantly (about four orders of
magnitude in Fig.~\ref{sinus_cut}) the resulting asymptotic photon number.
Whether other suitable field combinations enhance additionally the discovery
potential of the Schwinger effect by a secondary probe needs more realistic
modelling, including the back reaction.
Similar to the Sauter pulse (cf.\ inset in Fig.~\ref{sauter_cut}) the emission
is nearly isotropic, thus providing favorable observation conditions
perpendicular to the background field(s) and their generating (laser) beams.

\section{Summary}
We consider in leading order the photon emission accompanying the
process of shaking real electron-positron pairs off the vacuum by the
time-limited action of an external (spatially homogeneous) electric field.
In contrast to photon emission at all wavelengths off a plasma at nonzero
temperature (may it be an electron-positron plasma or a quark-gluon plasma),
where rates are accessible in various formalisms, the non-perturbative character
of pair creation due to the dynamical Schwinger process restricts us to the
consideration of the final state occupancies, both of $e^+e^-$-pairs and
photons.
Nevertheless, the found photon spectra uncover all wavelengths too.
Soft photons in the optical regime are produced amass and their abundancies can
even exceed the abundancy of $e^+e^-$ pairs in the sub critical region.
Such a feature provides a promising signal of the Schwinger process and
overcomes the unfavorably small number of residual $e^+e^-$ pairs.
The non linear amplification of the final photon yield by the superposition of
two fields with different scales is for photons similar to the known effect in
the residual pair sector, thus further enhancing the discovery potential of the
secondary photon probe which should be exploited at ELI-NP.\medskip

\noindent\textbf{Acknowledgements:}
The authors gratefully acknowledge inspiring discussions with H.~Gies,
F.~Karbstein, R.~Alkofer, D.~B.~Blaschke and C.~Greiner.
Many thanks go to S.~Smolyansky and A.~Panferov for common work on the plain
Schwinger process.
The fruitful collaboration with R.~Sauerbrey and T.~E.~Cowan within the HIBEF
project lead to the present investigation.

\appendix\onecolumngrid
\section{The photon spectrum}
The differential spectrum of single photons with momenta $\vec k$ summed over
polarizations $\lambda$ at time instant $t$ is defined by
\begin{align}
\frac{\diff^3N_\gamma(t,\vec k)}{\diff^3k} = \frac{1}{(2\pi)^3}\sum_\lambda
\langle0|a_{\lambda,H}^\dagger(t,\vec k)a_{\lambda,H}(t,\vec k)|0\rangle
\label{n_gamma}
\end{align}
where $a_{\lambda,H}^\dagger$/$a_{\lambda,H}$ are corresponding
creation/annihi\-lation operators in the Heisenberg picture ($H$);
in the interaction picture (I)\footnote{
For the reader's convenience we recall the transformation of operators $O$
between the various pictures.
The Heisenberg picture (H) follows from (i) the Schrödinger picture (S) by
$O_H(t) = U^\dagger(t,t_0)\, O_S(t_0)\, U(t,t_0)$ or from (ii) the interaction
picture (I) by $O_H(t) = U_\text{int}^\dagger(t,t_0)\, O_I(t)\,
U_\text{int}(t,t_0)$, and (I) from (iii) (S) by $O_I(t) = U_0^\dagger(t,t_0)\,
O_S(t_0)\,U_0(t,t_0)$;
(iii) causes $a_{\lambda,I}(t,\vec k) = U_0^\dagger(t,t_0)\, a_\lambda(\vec k)
\eh{-i\omega t_0}\, U_0(t,t_0) = a_\lambda(\vec k)\eh{-i\omega t}$ and (ii)
causes $a_{\lambda,H}(t,\vec k) = U_\text{int}^\dagger(t,t_0)\,
a_{\lambda,I}(t,\vec k)\, U_\text{int}(t,t_0) = U_\text{int}^\dagger(t,t_0)\,
a_\lambda(\vec k)\eh{-i\omega t}\, U_\text{int}(t,t_0)$.
}
the photon field operator $\mathcal A_I^\mu(t,\vec x)$ obeys the general
decomposition
\begin{align}
\mathcal A_I^\mu(t,\vec x) = \int\!\!\frac{\diff^3k}{\sqrt{2\omega}(2\pi)^3}
\sum_\lambda \Bigl[a_\lambda(\vec k)\varepsilon_\lambda^\mu(\vec k)\eh{-ikx}
+ a_\lambda^\dagger(\vec k){\varepsilon_\lambda^\mu}^*(\vec k)\eh{ikx}
\Bigr]
\end{align}
with $k^2 = \omega^2-\vec k^2 = 0$ and $\epsilon_\lambda^\mu(\vec k)k_\mu = 0$,
pointing to on-shell photons propagating on the light cone with two transverse
polarizations ($\mu$ is a Lorentz index).
The vacuum definition employed in~\eqref{n_gamma} reads $a_\lambda(\vec k)
|0\rangle = 0$ w.r.t.\ the photons;
the photons in turn are sourced by a Dirac current operator driving the photon
dynamics according to the wave equation
\begin{align}
\partial^2 \mathcal A_H^\mu(t,\vec x) = ej_H^\mu(t,\vec x)
\label{wave_eq}
\end{align}
with gauge conditions $\mathcal A_H^0 = 0$, $\vec\nabla\!\cdot\!
\vec{\mathcal A}_H = 0$ which are equivalent to $\epsilon_\lambda^0(\vec k) = 0$
and $\vec\epsilon_\lambda(\vec k) \!\cdot\!\vec k = 0$.
Equation~\eqref{wave_eq} is solved by a suitable unitary operator
$U_\text{int}(t,t_0)$ via $\mathcal A_H(t,\vec x) = U_\text{int}^\dagger
(t,t_0)\,\mathcal A_I(t,\vec x)\, U_\text{int}(t,t_0)$ and $j_H^\mu(t,\vec x) =
U_\text{int}^\dagger(t,t_0)\, j_I^\mu(\vec x)\,U_\text{int}(t,t_0)$, where the
current operator $j_I^\mu$ is constrained to $j_I^\mu(t,\vec x) =
{:}\bar\Psi_I(t,\vec x)\gamma^\mu \Psi_I(t,\vec x){:}$.
The notation ${:}\cdots{:}$ stands for normal ordering w.r.t.\ the vacuum
$|0\rangle$ and the operators $c_r$ and $d_r$ introduced below in~\protect\eqref
{psi_dec}.
This constraint omits the vacuum expectation value of $\bar\Psi_I\gamma^\mu
\Psi_I$, which is non-zero in a background field and creates a c-number
component of $\mathcal A_H$ which counteracts to the externally applied
background field $A$.
We neglect that backreaction (see e.g.~\cite{bloch_pair_1999}) since we are
interested here in the quantum part of the radiation field, which is henceforth
dealt with in the probe limit.

The needed Dirac wave operator can be decomposed in the interaction picture as
\begin{align}
\Psi_I(t,\vec x) = \int\!\!\frac{\diff^3p}{(2\pi)^3}\sum_r
\Bigl[c_r(\vec p)u_r(t,\vec p,\vec x)
+ d_r^\dagger(\vec p)v_r(t,\vec p,\vec x)\Bigr]
\label{psi_dec}
\end{align}
which extends the vacuum definition by $c_r|0\rangle = d_r|0\rangle = 0$;
$c_r$ and $d_r^\dagger$ carry the operator character and $u_r$ and $v_r$ the
bispinor structure.

In the interaction picture, the fermion dynamics obeys the Dirac equation
\begin{align}
\bigl\{i\gamma^\mu(\partial_\mu + ieA_\mu) + m\bigr\}\Psi_I(t,\vec x) = 0\:.
\label{dirac_eq}
\end{align}
We assume our purely electric background field $A_\mu$ to be spatially
homogeneous, but time dependent, which allows to split off the $\vec x$
dependence of the wave functions by replacing $u_r(t,\vec p,\vec x) \to
u_r(t,\vec p)\eh{i\vec p\vec x}$ and $v_r(t,\vec p,\vec x) \to v_r(t,\vec p)
\eh{-i\vec p\vec x}$ in~\eqref{psi_dec} with
\begin{align}
\bigl\{i\gamma^0\partial_t - \vec\gamma(\vec p-e\vec A(t)) -
m\bigr\}
u_r(t,\vec p) = 0\tag{\ref{dirac_eq}'}
\end{align}
(same for $v_r(t,-\vec p)$) and initial conditions $u_r(t\to-\infty,\vec p)
\propto u_r(\vec p) \eh{-i\sqrt{m^2+\vec p^2}\,t}$ and $v_r(t\to\infty,\vec p)
\propto v_r(\vec p) \eh{i\sqrt{m^2+\vec p^2}\,t}$.

With these ingredients we evaluate~\eqref{n_gamma} by employing
$a_{\lambda,H}(t,\vec k) = U_\text{int}^\dagger(t,t_0)\,
a_{\lambda,I}(t,\vec k)\, U_\text{int}(t,t_0)$ with Dyson's series
\begin{align}
U_\text{int}(t,t_0) = \Texp\biggl(\!-i\int_{t_0}^t\!\diff t' f_\varepsilon(t')
H_{\text{int},I}(t')\biggr)\cong 1-i\int_{t_0}^t\!\!\diff t'\,
f_\varepsilon(t') H_{\text{int},I}(t') + \mathrm O(e^2)\:,
\label{dyson}
\end{align}
where T means the time ordering operation and $f_\varepsilon(t) =
\eh{-\varepsilon|t|}$ is used to adiabatically turn the interaction on and off.
At the end of our calculation, we let $\varepsilon\to0$.
We restrict ourselves to the leading-order non-trivial term of~\eqref{dyson}
and utilize\footnote{\protect\parbox[t]{0.9\textwidth}{
We note the relations $H_I = H_{0,I} + H_{\text{int},I}$ with $H_{0,I} =
\int\!\diff^3x \bigl{\{}\overline\Psi_I\bigl[\vec\gamma\bigl(-i\vec\nabla -
e\vec A + m\bigr]\Psi_I + \frac12 \bigl[
\dot{\vec{\mathcal A}}_I^2 +
\bigl(\vec\nabla\times\vec{\mathcal A}_I\bigr)^2\bigr]
\bigr{\}}$.
}}
\begin{align}
H_{\text{int},I}(t) = e\int\!\!\diff^3x \mathcal A^\mu_I
(t,\vec x)j_{I,\mu}(t,\vec x)\:.
\end{align}
This yields for $a_{\lambda,H}$ up to order $\mathcal O(e^3)$
\begin{align}
\begin{aligned}
a_{\lambda,H}(t,\vec k) &= \left[1+i\int_{t_0}^t\!\!\diff t'H_{\text{int},I}(t')
\right] a_{\lambda,I}(t,\vec k)\left[1-i\int_{t_0}^t\!\!\diff t'H_{\text{int},I}
(t')\right]\\
&= \left[a_\lambda(\vec k) + i\int_{t_0}^t\!\!\diff t'\!\!\int\!\!\diff^3x\,
{:}\bar\Psi_I(t',\vec x)\left[e\gamma_\mu \mathcal A_I^\mu(t',\vec x),
a_\lambda(\vec k)\right]\Psi_I(t',\vec x){:}\right]\eh{-i\omega t}\\
&= \left[a_\lambda(\vec k) - i\frac{e\varepsilon_\lambda^{*\mu}(\vec k)}
{\sqrt{2\omega}} \int_{t_0}^t\!\!\diff t'\!\!\int\!\!\diff^3x\,
{:}\bar\Psi_I(t',\vec x)\gamma_\mu\Psi_I(t',\vec x){:}\eh{ikx'}\right]
\eh{-i\omega t}\:.
\end{aligned}
\end{align}
Insertion into~\eqref{n_gamma} lets us arrive at
\begin{align}
\begin{aligned}
&\frac{\diff^3N_\gamma(t)}{\diff^3k} = \frac{e^2}{(2\pi)^6}\frac{1}{2\omega}
\sum_\lambda\varepsilon_\lambda^\mu(\vec k)\varepsilon_\lambda^{*\nu}(\vec k)
\int_{t_0}^t\!\!\diff t_1\!\! \int_{t_0}^t\!\!\diff t_2\!\!
\int\!\!\diff^3x_1\!\! \int\!\!\diff^3x_2\,
f_\varepsilon(t_1)f_\varepsilon(t_2)\\
&\qquad\qquad\quad\times\langle0|
{:}\bar\Psi_I(t_1,\vec x_1)\gamma_\mu\Psi_I(t_1,\vec x_1){:}
{:}\bar\Psi_I(t_2,\vec x_2)\gamma_\nu\Psi_I(t_2,\vec x_2){:}
|0\rangle\eh{-ik(x_1-x_2)}\\
&= \frac{e^2}{(2\pi)^6}\frac{1}{2\omega} \sum_{\lambda,r,s}
\varepsilon_\lambda^\mu(\vec k)\varepsilon_\lambda^{*\nu}(\vec k)
\int\!\!\diff^3x \int\!\!\frac{\diff^3p}{(2\pi)^3}
\int_{t_0}^t\!\!\diff t_1 \bar v_r(t_1,-\vec p) \gamma_\mu
u_s(t_1,\vec p-\vec k) f_\varepsilon(t_1) \eh{-i\omega t_1}\\
&\qquad\times \int_{t_0}^t\!\!\diff t_2 \bar u_s(t_2,\vec p-\vec k) \gamma_\mu
v_r(t_2,-\vec p) f_\varepsilon(t_2) \eh{-i\omega t_2}\:.
\label{dn_reduced}
\end{aligned}
\end{align}
Note that $\diff^3N_\gamma(t=t_0)/\diff^3k = 0$.
We define the dimensionless photon phase-space occupation number
$f_\gamma(\vec k)=\diff^6N_\gamma(t\to\infty,\vec k)/\diff^3x\,\diff^3k$ and
get the basic equations~(\ref{n_from_C},\ref{C}).
We emphasize again that~(\ref{n_from_C},\ref{C}) are independent of a special
``driver'' of the dynamics of $u_r(t)$ and $v_r(t)$, e.g.\ omitting in the Dirac
equation the external field $A$ and allowing instead for a dynamical effective
mass $m(t)$, steered by the coupling to another background, one recovers the
results of~\cite{michler_asymptotic_2014}, albeit noted here in a different
form.

\section{Soft photons}
To study the soft photon limit one may split the time integral in Eq.~\eqref{C}
in the main text according to $\int_{-\infty}^\infty\diff t =
\int_{-\infty}^{-t_m}\diff t + \int_{-t_m}^{t_m}\diff t +
\int_{t_m}^\infty\diff t$, where $t_m$ stands for a matching scale with the
meaning that the background field $\vec A$ induces a noticeable dynamics of the
fermion field only within $-t_m\dots t_m$, that is $\dot{\vec A}(t\le t_m) =
\dot{\vec A}(t\ge t_m) = 0$.
We set $\vec A(t\le t_m) = 0$ and $\vec A(t\ge t_m) = \vec A_\infty$ and
elaborate $\lim_{\omega\to0} C_{rs\mu}$.
Employing
\begin{align}
\begin{aligned}
&\begin{alignedat}{2}
u_r(t\le-t_m,\vec p) &= \eh{-i\Omega(\vec p)(t+t_m)} &&u_r(\vec p)\:,\\
v_r(t\le-t_m,-\vec p) &= \eh{i\Omega(\vec p)(t+t_m)} &&v_r(-\vec p)\:,\\
\end{alignedat}\\
&\mkern14mu\begin{alignedat}{5}
u_r(t\ge t_m,\vec p) &={} &
&\alpha(t_m,\vec p) & &\eh{-i\Theta(t_m,\vec p)}
& &\eh{-i\Omega(\vec P_\infty)(t-t_m)} & &u_r(\vec P_\infty)\\
& &{}+{}&\beta(t_m,\vec p) & &\eh{i\Theta(t_m,\vec p)}
& &\eh{i\Omega(\vec P_\infty)(t-t_m)} & &v_r(-\vec P_\infty)\:,\\
v_r(t\ge t_m,-\vec p) &={} &
{}-{}&\beta^*(t_m,\vec p) &&\eh{-i\Theta(t_m,\vec p)}
& &\eh{-i\Omega(\vec P_\infty)(t-t_m)} & &u_r(\vec P_\infty)\\
& &{}+{}&\alpha^*(t_m,\vec p) & &\eh{i\Theta(t_m,\vec p)}
& &\eh{i\Omega(\vec P_\infty)(t-t_m)} & &v_r(-\vec P_\infty)
\end{alignedat}
\end{aligned}
\label{uv_asy}
\end{align}
with $\Omega(\vec p)^2 = m^2+\vec p^2$, $\Theta(t,\vec p) = \int_{-t_m}^t\!\!
\diff t'\,\Omega(\vec p-e\vec A(t'))$ and $\vec P_\infty = \vec p
- e\vec A_\infty$ from a Bogoliubov transformation (see below) results in the
leading order term
\begin{align}
\lim_{\omega\to0} C_{rs\mu}(\vec p,\vec k) = -i\alpha(t_m,\vec p)
\beta(t_m,\vec p)\Biggl[\frac{\bar v_r(-\vec P_\infty) \gamma_\mu
v_s(-\vec P_\infty)}{\omega+\frac{\vec P_\infty\vec k}{\Omega(\vec P_\infty)}}
+\frac{\bar u_r(\vec P_\infty) \gamma_\mu
u_s(\vec P_\infty)}{\omega-\frac{\vec P_\infty\vec k}{\Omega(\vec P_\infty)}}
\Biggr] + \mathcal O(\omega^0)\:.
\label{C_small_om}
\end{align}
The relation~\eqref{C_small_om} shows that $\lim_{\omega\to0}C \propto 1/\omega$
for a non-zero Bogoliubov coefficient $\beta(t_m,\vec p)$,
while $\lim_{\omega\to0}C$ (labels and index suppressed) remains finite for
$\beta(t_m,\vec p)=0$ due to the $\mathcal O(\omega^0)$ term.
As a consequence, in the former case $f_\gamma \propto 1/\omega^3$, while in the
latter case $f_\gamma \propto 1/\omega$.
$\beta(t_m,\vec p)\ne0$ implies an asymptotic pair density $N_{e^+e^-} \propto
|\beta|^2$, that is a specific soft photon spectrum accompanying a non-zero
residual pair number.
In the terminology of~\cite{michler_asymptotic_2014}, these contributions refer
to bremsstrahlung terms.
We emphasize here the mere use of well defined \emph{in}- and \emph{out}-states
and employed correspondingly a time-limited action of the background field.

In deriving~(\ref{uv_asy},\ref{C_small_om}) we use the Bogoliubov transformation
to solve the Dirac equation.
Introducing the Hamiltonian $h(\vec p) = \gamma^0(\vec p\vec\gamma + m)$ in
first quantization and the canonical momentum $\vec P(t) = \vec p-e\vec A(t)$
the governing equations for $u_r$ and $v_r$ read
\begin{alignat}{2}
\left\{i\partial_t - h\bigl(\vec P(t)\bigr)\right\}
u_r(t,\vec p) &= 0\:,\quad&
\left\{i\partial_t - h\bigl(\vec P(t)\bigr)\right\}
v_r(t,-\vec p) &= 0\:,\label{uv_dirac}\\
u_r(-t_m,\vec p) &= u_r(\vec p)\:,\quad& v_r(-t_m,-\vec p) &= v_r(-\vec p)\:.
\label{uv_init_cond}
\end{alignat}
We chose our initial condition at $t=-t_m$.
Since $\vec A$ points along the $z$-direction, $\vec A(t) = A(t)\vec e_z$, we
use an ansatz for $u_r(\vec p)$ and $v_r(-\vec p)$
\begin{align}
u_r(\vec p) = \frac{\Omega(\vec p) + h(\vec p)}{\sqrt{2\Omega(\vec p)
\bigl(\Omega(\vec p) - p_z\bigr)}}R_r\:,\qquad
v_r(-\vec p) = \frac{-\Omega(\vec p) + h(\vec p)}{\sqrt{2\Omega(\vec p)
\bigl(\Omega(\vec p) + p_z\bigr)}}R_r\:,
\end{align}
where $R_r$ denote two spinors ($r=1,2$) that are eigenvectors of
$\gamma^0\gamma^3$ with the eigenvalue $-1$.
With this ansatz, $u_r$ and $v_r$ are orthogonal and have the following
convenient properties:
\begin{alignat}{2}
h(\vec p) u_r(\vec p) &= \Omega(\vec p)u_r(\vec p)\:,\quad&
h(\vec p) v_r(-\vec p) &= -\Omega(\vec p)v_r(-\vec p)\:,\label{uv_eigen}\\
\partial_t u_r\bigl(\vec P(t)\bigr) &= \frac{eE(t)\eperp}
{2\Omega\bigl(\vec P(t)\bigr)^2}
v_r\bigl(-\vec P(t)\bigr)\:,\quad&
\partial_t v_r\bigl(-\vec P(t)\bigr) &= -\frac{eE(t)\eperp}
{2\Omega\bigl(\vec P(t)\bigr)^2}
u_r\bigl(\vec P(t)\bigr)\:,
\label{uv_can_der}
\end{alignat}
with $E(t)=-\dot A(t)$ the electric field and $\eperp = \sqrt{m^2+p_x^2+p_y^2}$
the transverse energy.
With these base spinors, the full solutions $u_r(t,\vec p)$ and $v_r(t,-\vec p)$
are seeked in the form
\begin{alignat}{5}
u_r(t,\vec p) &={}&
\alpha(t,\vec p) \eh{-i\Theta(t,\vec p)} &u_r\bigl(\vec P(t)\bigr) &&{}+{}&
\beta(t,\vec p) \eh{i\Theta(t,\vec p)} &v_r\bigl(-\vec P(t)\bigr)\:,
\label{u_ansatz}\\
v_r(t,-\vec p) &={}&
-\beta^*(t,\vec p) \eh{-i\Theta(t,\vec p)} &u_r\bigl(\vec P(t)\bigr) &&{}+{}&
\alpha^*(t,\vec p) \eh{i\Theta(t,\vec p)} &v_r\bigl(-\vec P(t)\bigr)\:,
\label{v_ansatz}
\end{alignat}
which directly lead to~\eqref{uv_asy}.
Plugging~\eqref{u_ansatz} together with~\eqref{uv_eigen} and~\eqref{uv_can_der}
into~\eqref{uv_dirac} leads to the following coupled equations for $\alpha$ and
$\beta$ (the ansatz~\eqref{v_ansatz} leads to the same equations):
\begin{alignat}{3}
\dot\alpha(t,\vec p) &={} &
&\frac{eE(t)\eperp}{2\Omega\bigl(\vec P(t)\bigr)^2}
\eh{2i\Theta(t,\vec p)} & &\beta(t,\vec p)\:,\\
\dot\beta(t,\vec p) &={} &
{}-{}&\frac{eE(t)\eperp}{2\Omega\bigl(\vec P(t)\bigr)^2}\,
\eh{-2i\Theta(t,\vec p)} & &\alpha(t,\vec p)\:,
\end{alignat}
which are solved numerically.
The initial conditions~\eqref{uv_init_cond} translate to $\alpha(t=-t_m,\vec p)
= 1$ and $\beta(t=-t_m,\vec p) = 0$.
The meaning of $\alpha$ and $\beta$ comes from $N_{e^+e^-}(t\to\infty,\vec p) =
2|\beta(t\to\infty,\vec p)|^2$, i.e.\ $\beta$ determines directly the number of
pairs created by the electric background field.

\bibliography{lit}
\end{document}